\documentstyle [psfig]{l-aa}
\begin{document}
\thesaurus{01(11.17.3; 12.04.1; 12.07.1)}
\title{Caustic Crossings in Quasar Light Curves?}
\author{M.R.S. Hawkins}
\institute{Royal Observatory,
Blackford Hill,
Edinburgh EH9 3HJ,
Scotland, UK}

\date{Received August 5; accepted October 21, 1998}

\maketitle

\begin{abstract}

Numerical simulations and theoretical studies of the gravitational
microlensing effect of a population of small bodies distributed along
the line of sight to a compact light source such as a quasar indicate
that caustic crossing effects will be present given a sufficiently
large optical depth to lensing.  These events will produce
characteristic patterns in the quasar light curve.  In this paper we
use a large sample of quasar light curves to search for features
with the properties of caustic crossing events.  Two good candidates
are presented which it is argued are not easily explained as intrinsic
variation of the quasars.  The relation between these events and
microlensing features seen in multiple quasar systems is discussed,
as well as the implications for the dark matter problem of a
population of microlensing bodies.


\end{abstract}

\section{Introduction}

Over the past 10 years or so several groups
have published numerical simulations of the gravitational microlensing
effect of a distribution of compact bodies along the line of sight
to a distant light source such as a quasar.  These light curves
contain a number of identifiable features which depend on the optical
depth to lensing, the size and structure of the source and the mass and
velocity of the lenses.  In particular, lenses can combine non-linearly
to produce caustics which result in characteristic high-amplification
events.  In this paper we make the case that such features may
actually be seen in observed quasar light curves, and present examples
from a large scale monitoring programme which have all the
properties of caustic crossing events.  This implies the existence
of a large population of microlensing bodies with masses which can be
estimated from the duration of the caustic crossings to be subsolar.

The first numerical simulations of microlensing (Paczy\'{n}ski 1986)
show undulating variation, punctuated by double spiked structures
produced by caustic crossings.  Since then a number of more
sophisticated simulations have been carried out (Kayser et al. 1986,
Schneider \& Weiss 1987, Lewis et al. 1993), as well as studies of the
effect of microlensing on extended sources (Refsdal \& Stabell 1991).
The question of whether microlensing is actually seen in quasar light
curves has also been investigated (Schneider \& Weiss 1987), and it
seems clear that, at least in the case of multiply lensed quasars,
microlensing is taking place (Schild 1996, Irwin et al. 1989, Corrigan
et al. 1991).  The possibility that all quasars at sufficiently high
redshift are being microlensed, and that this accounts for most of the
observed variation, has been argued in a number of papers (Hawkins 1993,
1996, Hawkins \& Taylor 1997) on the basis of statistical analysis of
the quasar light curves.  The relevance of most numerical simulations
to observations of quasar light curves depends on the optical depth
for microlensing of any population of lenses.  If they have an
optical depth around unity as they will in multiply lensed systems
then complicated caustic patterns can be expected, and this will also
be the case for a cosmological critical density of lenses.  However
the nature of the variation is different for low optical depth where
individual events can be distinguished, as illustrated by Schneider
(1993).\\

Statistical tests are inevitably blunt weapons, and there is much to be
learned from examining the structure of individual light curves with a
view to identifying features which can unambiguously be attributed to
microlensing.  In this paper light curves for a large sample of
isolated quasars are examined for the characteristic features of
caustic crossings produced by a large optical depth of microlenses.
Two good candidates are presented, and discussed on the basis of
current quasar models.\\

\section{Data analysis}

The parent sample of quasar light curves which was used for the
investigation has already been described in detail (Hawkins 1996,
Hawkins \& V\'{e}ron 1995), although it has been updated by the
addition of five more yearly epochs.  It is based on COSMOS and
SuperCOSMOS measures of a large set of plates taken with the UK Schmidt
telescope in ESO/SERC Field 287 centered on 21h 38m, -45$^{\circ}$
(1950).  The dataset contains homogeneous yearly coverage from 1983 to
1997 in two colours, a blue passband (IIIa-J/GG395) and a red passband
(IIIa-F/RG630).\\

The idea of the project was to identify caustic crossing events in
the set of light curves.  The possibility of implementing an algorithm
for automatically detecting the characteristic twin spiked shape was
reluctantly rejected due to the difficulty of defining a sufficiently
general template.  The search was thus carried out by eye, which 
although exceptionally well suited to detecting patterns, results in
an inevitable loss of statistical objectivity.\\

The first impression on scanning through large numbers of light curves
in two colours is that the variation is on the whole close to being
achromatic, and this can be supported by a statistical test (Hawkins
1996).  A good example of such a light curve is shown in the top panel
of Fig. 1 where no systematic colour change occurs over a change in
brightness of 1.2 magnitudes.  More careful examination however
shows small departures from achromaticity for many of the light
curves, and in a few cases very large differences.
\begin{figure}
\psfig{figure=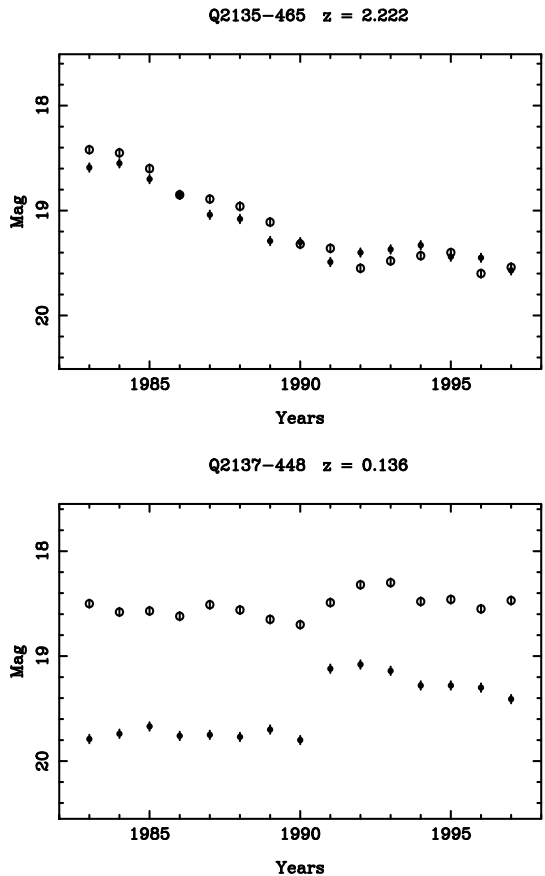,height=12cm,angle=0}
\caption[]{
      Light curves for two quasars.  Filled and open circles are
      blue and red passband measures respectively. Error bars are
      based on measured photometric errors.
}
\end{figure}
The bottom panel shows the light curve for a low luminosity low
redshift quasar undergoing a small outburst.  This feature is most
unlikely to be caused by microlensing due to the low redshift of
the quasar, but is very much what one would expect from an intrinsic
event.  The idea would be that an outburst in the hot blue core
of an accretion disk would propagate outwards becoming smoothed out
and degraded by the time it reached the cool red outer part of the
disk.  If there is a time lag between the two colours for the onset
of the event it is clearly less than a year.\\

Fig. 2 shows an event which appears to have the characteristics of
a caustic crossing.  The morphology of such events is well illustrated
by the various groups which have carried out numerical simulations
of microlensing (Kayser et al. 1986, Schneider \& Weiss 1987,
Lewis et al. 1993).  The double spiked features
are readily spotted in the set of light curves, but more often than
not are overlaid by short term events which distort the morphology.
The light curve in Fig. 2 is a particularly clean example.  The
variation is characterised by a rise in the blue to a double cusp
shaped feature followed by a fall.  The cusps are smoothed out in the
\begin{figure}[t]
\psfig{figure=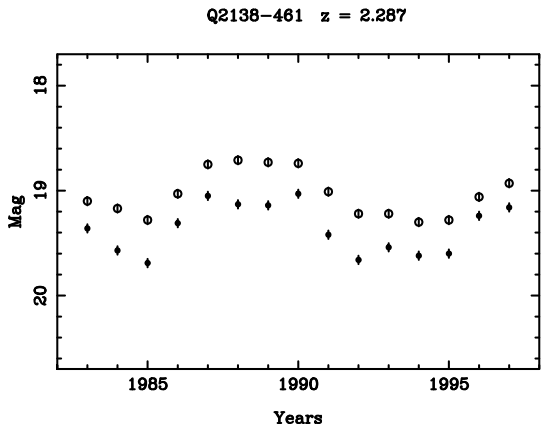,height=12cm,angle=0}
\vspace*{-6cm}
\caption[]{
 Light curve for a quasar showing a possible caustic crossing.
      Symbols as for Fig. 1
}
\end{figure}
red light curve which is symmetrically situated over the blue, with a
smaller amplitude.  The explanation for this in the context of
microlensing would be that a compact blue core is strongly amplified
by a lens with significantly larger Einstein radius during a caustic
crossing.  The red light curve however is dominated by the outer
parts of the accretion disk, which is of comparable size to the
Einstein radii of the lenses, and is amplified less strongly with
no discernible cusps.  This chromatic type of behaviour has been
discussed from a theoretical point of view by Wambsganss \&
Paczy\'{n}ski (1991).  It seems possible that one could contrive
an explanation for this light curve based on intrinsic variation,
but the symmetry of the configuration is hard to account for.  Also,
the cusp like features which are so much a feature of microlensing
have no natural explanation as intrinsic events.\\

Although the light curve in Fig. 2 is suggestive of a microlensing
event, the rather small amplitude leaves open the possibility that
it consists of a juxtaposition of small intrinsic events.
To circumvent this difficulty a detailed search was carried out
on the largest amplitude quasars ($\delta m > 1.5$) to look for
features with the characteristics of caustic crossings.  Of the
10 quasars with complete light curves in $B_{J}$ and $R$, the
two best candidates are shown in Fig. 3.  In each case the blue
flux rises and falls sharply, within the space of about a year.
\begin{figure}
\psfig{figure=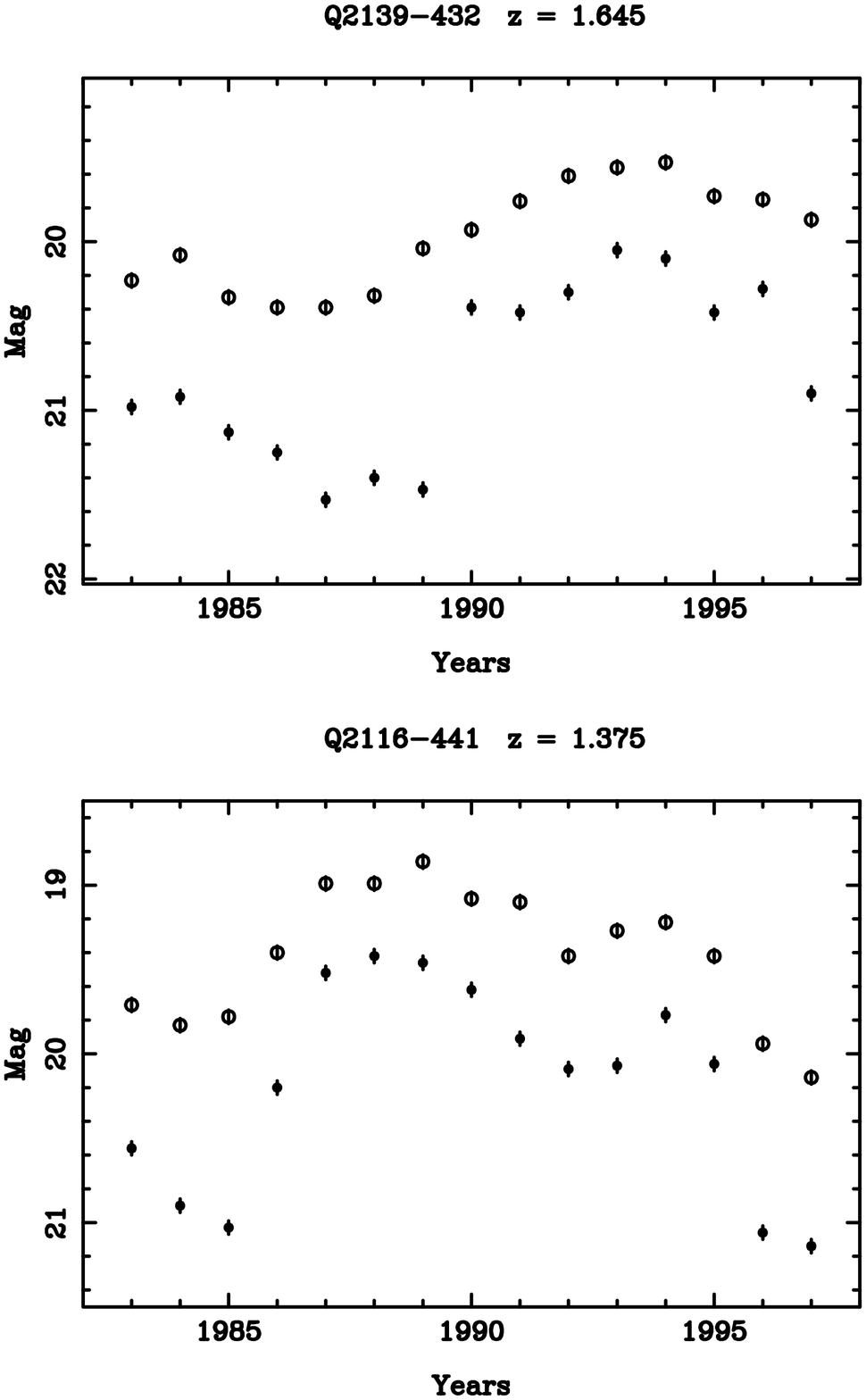,height=12cm,angle=0}
\caption[]{
      Light curves for two quasars showing all the characteristics
      expected of caustic crossing events.  Symbols as for Fig. 1.
}
\end{figure}
The red flux on the other hand actually starts to increase two years
or so before the blue, and appears to continue to fall after the
blue has bottomed out.  It does however only achieve a much smaller
amplitude.  This is very hard to understand in the context of intrinsic
variation, but as before is in accord with the microlensing model.
In this case, the larger size of the accretion disk in the red produces
early amplification of the source, but this same large size results in
a smaller total amplitude than in the blue.\\ 

\section{Discussion and conclusions}

The availability of light curves for the components of multiply
lensed quasars can in principle provide unambiguous templates
for intrinsic variation and microlensing.  The best data available
at present is for the double quasar Q0957+561.  Intensive CCD
monitoring by two groups (Schild \& Thomson 1997, Kundi\'{c} et
al. 1997) from 1993 to 1996 show small fluctuations of around 0.15
magnitudes which are very well matched in both images after
allowing for the time delay.  These must be intrinsic variations
and appear to be similar in character, although of a smaller
amplitude, to those seen in the Seyfert
galaxy NGC 5548 by Clavel et al. (1991).  This similarity extends to
chromatic effects.  Kundi\'{c} et al. (1997) show that when Q0957+561
gets intrinsically brighter it becomes bluer, as seen in NGC 5548.
It is only when monitored over much longer timescales that
convincing microlensing effects become apparent (Schild \& Thomson
1993).  When the light curves of the two components are subtracted
after applying the time delay, the dominant feature is a long term
decrease in the magnitude difference of about 0.3 magnitudes over
a period of 12 years.  Effects of this sort are seen in the Field
287 quasar sample, but do not appear to be associated with a caustic
crossing.\\

The time interval between the peaks provides a very rough way of
estimating the mass of the microlensing bodies.  In simple two mass
systems the separation of the caustics is typically a few tenths of
an Einstein radius (Schneider \& Weiss 1986).  For more complex
arrangements of lenses the separations can be larger, up to a few
Einstein radii (Lewis et al. 1993).  The light curves in Figs 2 and 3
have separations of around four years between the caustics, and if we
adopt a typical lens velocity across the line of sight of 600
km.sec$^{-1}$, the distance travelled is around $2 \times 10^{-3}$ pc.
This implies (Hawkins 1996) a lens mass of $10^{-2}M_{\odot}$, a
factor of ten larger than that derived from a statistical treatment of
the light curves (Hawkins 1996).  The uncertainties in this calculation
are rather large, at least an order of magnitude, but if this
difference is real it is probably due to the fact that it is the most
massive lenses which are likely to give the most pronounced caustic
patterns.\\

The relative ease with which events resembling caustic crossings
can be found in the set of light curves strongly suggests that they
could be a fundamental aspect of the variation mechanism.  It is
clearly not possible to rule out any conceivable intrinsic mechanism
which can produce the observed features, but until such a mechanism is
proposed it seems a viable alternative to attribute them to
microlensing.  This then adds to the existing case for a population of
planetary mass bodies sufficient to account for the dark matter.
Other lines of argument include the statistical analysis of quasar light
curves (Hawkins 1996), microlensing in gravitationally lensed multiple
quasar systems (Schild 1996, Hawkins 1997) and the recent detection
of a Jupiter mass body by the MACHO team (Bennett et al. 1997,
Hawkins 1998).  Although none of these strands of evidence is
conclusive in itself, the case for dark matter in the form of
planetary mass compact bodies is becoming steadily more broadly
based.\\

\end{document}